\documentclass[doublecol]{epl2}
\usepackage{amsmath,amssymb,amsthm,graphicx,latexsym}

\newcommand{\bd}{\begin{document}}
\newcommand{\ed}{\end{document}}
\newcommand{\bc}{\begin{center}}
\newcommand{\ec}{\end{center}}
\newcommand{\vs}{\vspace}

\begin{document}

\title{Dissipative nonlinear waves in a gravitating quantum fluid }

\author{Biswajit Sahu\inst{1}\thanks{ e-mail: bisu.sahu@gmail.com}, Anjana Sinha\inst{2}\thanks{ e-mail: anjana23@rediffmail.com; sinha.anjana@gmail.com} and R. Roychoudhury\inst{3,4}\thanks{e-mail : rajdaju@rediffmail.com; rroychoudhury123@gmail.com - fax : +91 33 24146321; phone : +91 33 25753020}}

\abstract{Nonlinear wave propagation is studied analytically in a dissipative, self-gravitating Bose Einstein condensate,
in the framework of Gross-Pitaevskii model. The linear dispersion relation shows that the effect of dissipation is to suppress dynamical instabilities that destabilize the system. The small amplitude analysis using reductive perturbation technique is found to yield a modified
form of KdV equation. The soliton energy, amplitude and velocity are found to decay with time, whereas the soliton width increases, such that the
soliton exists for a finite time only. 
}


\institute{
\inst{1} Department of Mathematics, West Bengal State University, Barasat, Kolkata - 700 126, INDIA \\
\inst{2} Physics and Applied Mathematics Unit, Indian Statistical Institute, Kolkata - 700 108, INDIA \\
\inst{3} Department of Mathematics, Bethune College, 181, Bidhan Sarani, Kolkata - 700 006, INDIA \\
\inst{4} Advanced Centre for Nonlinear and Complex Phenomena, 1175
Survey Park,  Kolkata - 700075, INDIA}

\maketitle

\section{1 Introduction}

Theoretically predicted by Bose and Einstein in 1925
\cite{chavanis1} and first produced experimentally by the Cornell
and Wiemann group some seven decades later in 1995 \cite{cornell},
Bose Einstein Condensates (BECs) have fascinated scientists for
their intriguing properties \cite{bec2,bec3,bec4,bec5,bec6,bec7,proca}.
The major role they play in condensed matter physics is pretty
well known \cite{cond}. Fairly recent studies suggest that self
gravitating BECs could play an important role in astrophysics and
cosmology, in the study of neutron stars and dark matter halos
\cite{chavanis2,shapiro}. It is anticipated that a substantial amount of
matter in neutron stars might exist as BECs due to their
superfluid quantum core. However, such compact astronomical
objects have very strong magnetic fields and will be considered in
a future work. The existence of dark matter might be linked to the
observed flat rotation curves of galaxies. Dark matter halos could
be gigantic quantum objects, interpreted as self-gravitating BECs
at $T=0$. At large scales, quantum effects are negligible and the
classical hydrodynamic equations are good enough in explaining the
large-scale structure of the universe. However, at small-scales,
the pressure arising from the Heisenberg uncertainty principle or
from the repulsive scattering of the bosons may stabilize dark
matter halos against gravitational collapse. In case the bosons
are assumed to be non self-interacting, gravitational collapse is
prevented by the Heisenberg Uncertainty Principle, which is
equivalent to a quantum pressure. In such cases the mass of the
individual bosons $ \approx 10^{-24}$ eV/c$^2$. If the bosons are
assumed to have a repulsive self interaction, gravitational
collapse is prevented by the pressure arising from the scattering.
For scattering lengths $ \sim 10^6 $ fm, the mass of such bosons $
\approx 1 $ eV/c$^2$. This stabilizing of dark matter halos
against gravitational collapse
--- either due to pressure arising from Heisenberg Uncertainty
Principle, or the repulsive scattering of bosons --- could lead to
smooth core densities in agreement with observations \cite{smooth-core}, instead of
cuspy density profiles \cite{cuspy} predicted by the cold dark matter model \cite{cdm1,cdm2}.
In \cite{stenflo}, the authors studied Jeans instability in a self-gravitating dusty plasma,
considering an attractive force between two equally charged dust particles.
For attractive self interacting BECs with negative scattering
length, the ultimate collapse cannot be prevented, and the BEC is
not stable. This leads to the formation of supermassive black
holes at the centre of galaxies. Additionally, in the
astrophysical context, boson stars are conceptualized as
self-gravitating bosons, exclusively trapped in their own
gravitational potential \cite{ruffini}. Theoretically, boson stars
are objects of highly diverse sizes and masses, depending on the
assumed particle mass and the strength of self-interaction. For
example, miniboson stars are very compact objects with radii
$\approx 10 ^{-15} m$ \cite{lee-pang}, whereas those of the size
of the sun, but having mass $\sim 10 ^6$ solar masses, could mimic
supermassive black holes \cite{schunck-milke}. However, for such
compact bodies, we must use general relativity and couple the
Klein-Gordon equation to the Einstein field equations; hence these
are exempted from the present study.

With the experimental verification of BECs in magnetically trapped
dilute vapors of alkali metals \cite{cornell,alkali2}, interest in this particular field
has been revived in recent times \cite{nature1,nature2,nature3,prl107}. It has been
observed that dissipation plays a crucial role in the
stabilization of a BEC droplet by suppressing the dynamical
instabilities \cite{pra70} --- observation of collective damped
oscillations of the condensate shows the presence of dissipation
\cite{jin,mewes}. Additionally, the observation of vortex lattices
when the  trapping potential is rotated fast enough, implies the
presence of dissipation \cite{pra65}. In the present study, we shall
investigate the nonlinear wave propagation in a self gravitating
BEC, in the framework of Gross-Pitaevskii equation, in the
presence of dissipation. It is worth mentioning here that Ghosh
and Chakrabarti studied this problem in \cite{samiran-pre84}, but
without the dissipative term. They showed that the small amplitude
analysis is governed by the KdV equation. Our aim in this work is
to emphasize on the effect of the dissipative term on the
propagation of nonlinear structures.

This article is organized as follows. To make the paper self
contained, the basic equations are introduced in Section 2.
Section 3 is kept for the derivation and discussion of the linear
dispersion relation. The small amplitude analysis is carried out
in Section 4, using the standard reductive perturbation technique.
The nonlinear evolution is also derived analytically in Section 4.
Finally, Section 5 is kept for Discussions  and Conclusions.

\section{2 Basic Governing Equations}

Self gravitating quantum systems, proposed by Penrose in his
discussion of quantum state reduction by gravity \cite{penrose},
are a dilute quantum gas with short-range dipole interactions
between atoms. At very low temperatures ($T \rightarrow 0$), all
particles (bosons) in a dilute Bose gas confined in a trapping
potential $V_{ext}(\mathbf{r})$, condense to the same quantum
ground state and form a Bose Einstein condensate. The condensation
of the bosons takes place when their thermal (de Broglie)
wavelength $\lambda _{DB}= \left( 2 \pi \hbar ^2 / m k_B T \right)
^{1/2}  $ exceeds their mean separation. The system is described
by the one parameter condensate wave function $\psi
(\mathbf{r},t)$. We consider a system of $n$ bosons, with mass $m$
in interaction. For ultra cold temperatures, the dynamical
behaviour of a weakly interacting Bose gas in the presence of
dissipation, considering the mean field analysis, is described by
the Gross-Pitaevskii (GP) equation \cite{g-p}
\begin{equation}\label{gp}
    \displaystyle \left( i - \nu \right) \hbar \frac{\partial \psi}{\partial t}
    = \left[ - \frac{\hbar ^2}{2m} \nabla ^2 + V _{ext} (\mathbf{r}) + g |\psi | ^2 \right] \psi
\end{equation}
where $g = 4 \pi a_s \hbar ^2 / m$ is the interacting constant,
$a_s$ is the $s$-wave scattering length ($a_s >0$ for repulsive
self-interactions), and $\nu$ is a phenomenological dissipation
constant, the dissipation being caused by the interaction between
a BEC and thermal cloud \cite{xiao-xian}. Here $g | \psi | ^2$ is the
self-interaction term and $ n = \int dr | \psi (\mathbf{r},t) | ^2
$ denotes the total number of bosons in the condensate. It is
worth mentioning here that for BEC to take place, the $s$-wave
scattering length must be very small as compared to the de Broglie
wavelength : $\lambda _{DB} >> a_s$. However, in the presence of a
gravitational trap, no additional trap is required, and the external
trapping potential $V_{ext}$ may be replaced by $V_G$,
where $V_G = m \Phi $, $\Phi $ being the gravitational potential
obeying the Poisson equation
\begin{equation}\label{poisson}
    \displaystyle \nabla ^2 \Phi = 4 \pi G \left( \rho - \rho _0 \right)
\end{equation}
Here, $\rho = m n | \psi (\mathbf{r},t) | ^2$ is
the mass density, $\rho _0$ is its equilibrium value, and $n$ is
the number density. \\
In such a situation, the condition for stable BEC is : \\
$a_* >> \lambda _{DB} >> a_s , \ \ a_*
= 4 \pi ^2 \hbar ^2 / m u $, $a_*$ being the Bohr radius
associated with gravitational coupling $u$. \\
Typical characteristic values of the above parameters in a BEC are : $a_* \sim 10 $ cm, $\lambda _{DB} \sim 10
^{-5} - 10^{-3}$ m, $a_s \sim 3$ nm, $n = 10^{21} $ m$^{-3}$, and $m= 35.2
\times 10^{-27} $ kg \cite{samiran-pre84, physical}. \\

To rewrite the GP equation in the form of hydrodynamic equations, we apply the Madelung transformation
\begin{equation}\label{madelung}
    \psi (\mathbf{r},t) = \displaystyle \sqrt{\frac{\rho (\mathbf{r},t) }{m}} \exp [i S (\mathbf{r},t) / \hbar]
\end{equation}
where $S(\mathbf{r},t) $ has the dimension of action, and define the irrotational flow velocity as
\begin{equation}\label{v}
    \mathbf{v} = \mathbf{\nabla} \mathbf{S} /m
\end{equation}
The flow is irrotational since $\mathbf{\nabla} \times \mathbf{v} = 0$. \\

Substituting eq. (\ref{madelung}) in eq. (\ref{gp}), and separating
real and imaginary parts, we obtain
\begin{equation}\label{cont}
    \displaystyle \frac{\partial \rho }{\partial t} + \mathbf{\nabla} \cdot  (\rho \mathbf{v}) - \frac{2 \rho \nu}{\hbar} \frac{\partial \mathbf{S}}{\partial t} = 0
\end{equation}
which is equivalent to the continuity equation, and
\begin{equation}\label{mom}
\begin{array}{lll}
    \displaystyle \rho \left[ \frac{\partial \mathbf{v}}{\partial t} + \left( \mathbf{v} \cdot \mathbf{\nabla} \right) \mathbf{v} \right] &=&
    \displaystyle - \mathbf{\nabla} P - \frac{\rho}{m} \mathbf{\nabla} V_G \\ \\
    & & \displaystyle + \frac{\hbar ^2 \rho}{2 m^2} \mathbf{\nabla}
    \left( \frac{1}{\sqrt{\rho}} \nabla ^2 \sqrt{\rho} \right) \\ \\
    & & \displaystyle - \frac{\hbar \nu \rho}{2m} \mathbf{\nabla}
    \left( \frac{1}{ \rho } \frac{\partial \rho}{\partial t} \right) \\
    \end{array}
\end{equation}
which is equivalent to the momentum equation of hydrodynamics.
Here $P$ is the pressure term, and the term containing $\hbar ^2$
on the rhs is the quantum potential.

In BEC the non thermal pressure term $P$ arises from the short-range interactions
between the bosons, and is related to the mass density through
\cite{cond}
\begin{equation}\label{p}
    P = \displaystyle \frac{1}{2} g \left( \frac{\rho}{m} \right) ^2
\end{equation}

\textbf{The continuity equation (\ref{cont})can be written as}
\begin{equation}\label{cont-a}
    \displaystyle \frac{\partial \mathbf{S}}{\partial t} = \displaystyle \frac{\hbar}{2 \rho \nu}
    \left\{ \frac{\partial \rho }{\partial t} + \mathbf{\nabla} \cdot (\rho \mathbf{v}) \right\}
\end{equation}
\textbf{Taking gradient of equation (\ref{cont-a}), and using equation (\ref{v}), we obtain }
\begin{equation}\label{cont-1}
    \displaystyle m \frac{\partial \mathbf{v}}{\partial t} = \frac{\hbar}{2 \nu} \mathbf{\nabla} \cdot \left[
    \frac{1}{\rho} \left\{ \frac{\partial \rho }{\partial t} + \mathbf{\nabla} \cdot (\rho \mathbf{v}) \right\} \right]
\end{equation}
\textbf{Similarly, using equations (\ref{poisson}) and (\ref{p}), alongwith $V_G = m \Phi$, the momentum equation (\ref{mom})  reduces to :}
\begin{equation}\label{mom-1}
\begin{array}{lll}
    \displaystyle & & \displaystyle \mathbf{\nabla} \cdot \left[ \frac{\partial \mathbf{v}}{\partial t} + \left( \mathbf{v}
    \cdot \mathbf{\nabla} \right) \mathbf{v} \right]
    + \displaystyle  \frac{\hbar \nu }{2m} \nabla ^2 \left( \frac{1}{ \rho } \frac{\partial \rho}{\partial t} \right) \\ \\
    & & \displaystyle - \frac{\hbar ^2 }{2 m^2} \nabla ^2 \left( \frac{1}{\sqrt{\rho}}
    \nabla ^2 \sqrt{\rho} \right)  + \frac{c_s ^2  }{\rho _0} \nabla ^2 \rho \\ \\
    & & \displaystyle = - 4 \pi G \left( \rho - \rho _0 \right) \\
    \end{array}
\end{equation}
where $c_s = \sqrt{g \rho _0 / m^2} = \sqrt{gn/m} $ is the velocity of sound in the medium. For BEC comprising of sodium atoms, $c_s \approx 10^{-3} $ m/s \cite{physical}. \textbf{Equations (\ref{cont-1}) and (\ref{mom-1}) are the final set of equations which need to be analyzed simultaneously. In the absence of an exact analytical solution, numerical results and approximate methods like the linear dispersion relation, give a good estimate of the system behaviour. }

\section{3 Linear Dispersion Relation}

The linear dispersion relation is a powerful tool to give a
qualitative picture of dissipation in the system. To derive its
form, we shall start with  the equations (\ref{cont-1}) and
(\ref{mom-1}), and linearize them about equilibrium values $\mathbf{v}=0$
and $\rho (\mathbf{r},t)= \rho _0$,
with the perturbations varying as $\displaystyle e^{i \left( \mathbf{k} \cdot \mathbf{r} - \omega t \right) } $ \\
For brevity, we just quote the final result, without going into
the mathematical details. The linear dispersion relation is
obtained as a quadratic equation in $\omega $ :
\begin{equation}\label{dispersion}
\begin{array}{lll}
    & & \displaystyle \left( 1 + \nu ^2 \right) k^2 \omega ^2 +
\displaystyle i \ \frac{\nu}{\hbar} \ \left\{ 2 m c_s ^2
    \left(k^2 - k_J ^2 \right) + \frac{\hbar ^2 k^4}{m} \right\}  \omega \\ \\
    & & - \displaystyle \left\{c_s ^2 k^2 \left( k^2 - k_J ^2 \right) + \frac{\hbar ^2 k^6}{4 m^2} \right\}= 0 \\
    \end{array}
\end{equation}
where \textbf{$k=|\mathbf{k}|$,} $\omega _J = \sqrt{4 \pi G \rho _0} $ is the Jeans frequency
and $k_J = \omega _J / c_s $ is the Jeans wave number. Hence
\begin{equation}\label{omega}
    \omega = \displaystyle \pm \omega _R  + i \omega _I
\end{equation}
with
\begin{equation}\label{omega-i}
    \omega _I = \displaystyle  - \nu \ \frac{ m \displaystyle  \left\{ c_s ^2
    \left(k^2 - k_J ^2 \right) + \frac{\hbar ^2 k^4}{2m^2}
    \right\}}{ \displaystyle \left( 1 + \nu ^2 \right) \hbar k^2}
\end{equation}
and
\begin{equation}\label{omega-r}
    \omega _R = \displaystyle \frac{ \displaystyle \sqrt{ \left\{ \frac{\hbar ^2 k^8 }{m^2}
    + 4 c_s ^2 k^4 \left(k^2 - k_J ^2 \right) - \frac{4 m^2 \nu ^2}{\hbar ^2
    }c_s ^4 (k^2 - k_J ^2 ) ^2  \right\}}}{ \displaystyle 2 \left( 1 + \nu ^2 \right) k^2}
\end{equation}

\begin{figure}
 \begin{center}
 \includegraphics[width = 6 cm]{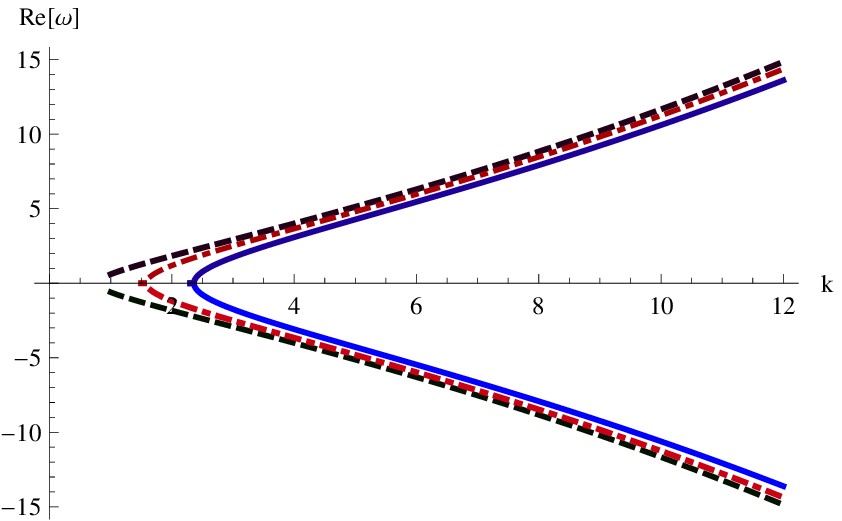} ~~~ \includegraphics[width = 6 cm]{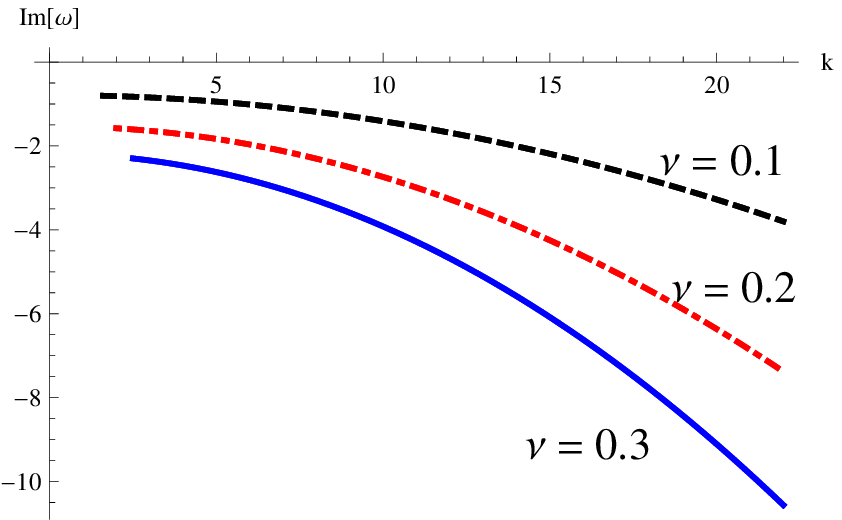} \\
  \caption{Plot of $ \omega _R$ and $\omega _I$ against $k$, for different values of the dissipation parameter $\nu$. While the real part of $\omega$ is more or less unaffected by the dissipation term, increasing the dissipation increases its imaginary part, and the nonlinear structure gets more and more damped. }
 \end{center}
 \end{figure}

In the absence of dissipation ($\nu = 0$), we get back the
dispersion relation given in \cite{samiran-pre84}
\begin{equation}\label{omega-0}
    \omega = \displaystyle \pm c_s \sqrt{(k^2 - k_J ^2 ) + H_Q ^2 \lambda ^2 k^4}
\end{equation}
where $H_Q = \hbar / (2 m c_s \lambda )$ is a dimensionless
quantum parameter, and $\lambda = 2 \pi / k $ is the perturbation
wavelength. Thus, in the absence of dissipation, the system is
stable only if the perturbation wave number is greater than the
Jeans wave number : $k > k_J$ or $\lambda < \lambda _J$. However,
in the presence of dissipation, $\omega $ may have an imaginary
part even for $k < k_J$ as observed from equation (\ref{omega-r}).
But, it is worth noting here that dissipation alone is responsible
for the imaginary part in $\omega $. For positive dissipation
constant $\nu$, $\omega _I$ is negative, showing a decaying wave.
Thus, analogous to the observation made in ref. \cite{pra70},
applying numerical simulation, our analytical approach arrives at
a similar result
--- the effect of dissipation is to resist dynamical
instabilities that destabilize the system. We plot the linear
dispersion relation for a dissipative gravitating fluid in Fig. 1.
For fixed value of the quantum parameter $H_Q$, the real part of
$\omega$ is more or less unaffected by the dissipation term $\nu$,
but the magnitude of its imaginary part increases with increasing
dissipation, and hence the system gets more and more damped. This
is depicted in Fig. 1. It is to be noted that $\omega$ has an
imaginary part only in the presence of dissipative term $\nu$.

\section{4 Evolution of Small Nonlinear Structures}

\vspace{.5cm}

\par
\textbf{We now consider a 3-dimensional Bose gas, confined tightly in two spatial dimensions, and weakly propagating in the $x$ direction only. Reduced dimensionality leads to strong quantum fluctuations which tend to destroy macroscopic coherence. However, since we are considering extremely low temperatures ($T \approx 0$), this effect can be neglected. Thus, in the limit of one spatial dimension, the longitudinal dynamics dominate the system, and the transverse modes are heavily suppressed. In this approximation, the wave function $\psi (\mathbf{r}, t)$ can be decoupled into the product of a time-independent transverse component (which does not contribute to the dynamics of the system in the present case) and a time dependent axial component $\psi (x,t)$. At temperatures close to zero where phase fluctuations are negligible such
weakly-interacting 1D and 2D condensates are possible, and have been studied theoretically as well as experimentally \cite{low-dim}. To explore the evolution of weak nonlinear structures in one spatial dimension,} for $k_J \ll k$, we shall apply the reductive perturbation technique \cite{rpt}. Before proceeding further, we write down the normalizations used:
\begin{equation}\label{normalize}
    \bar{x} = \displaystyle \frac{x}{\lambda} \ , \  \bar{t} = \displaystyle \frac{c_s t}{\lambda} \ , \
    \bar{\rho} = \displaystyle \frac{\rho}{\rho _0} \ , \ \bar{v} = \displaystyle \frac{v}{c_s}
\end{equation}
In terms of the new normalized variables, equations (\ref{cont-1}) and (\ref{mom-1}) take the forms
\begin{equation}\label{cont-2}
    \displaystyle \frac{\partial v}{\partial t} = \frac{H_Q}{\nu } \frac{\partial }{\partial x} \left[
    \frac{1}{\rho} \left\{ \frac{\partial \rho }{\partial t} + \frac{\partial}{\partial x}  (\rho v) \right\} \right]
\end{equation}
and
\begin{equation}\label{mom-2}
\begin{array}{lll}
    & & \displaystyle \frac{\partial}{\partial x} \ \left[ \frac{\partial v}{\partial t} + v \frac{\partial v}{\partial x} \right]
    + \nu H_Q \frac{\partial ^2}{\partial x ^2} \left( \frac{1}{ \rho } \frac{\partial \rho}{\partial t} \right) \\ \\
    & & \displaystyle - 2 H_Q ^2  \frac{\partial ^2}{\partial x ^2} \left( \frac{1}{\sqrt{\rho}} \frac{\partial ^2 \sqrt{\rho}}{\partial x^2} \right) + \frac{\partial ^2 \rho}{\partial x^2} \\ \\
    & & \displaystyle = - 4 \pi ^2 \left( \frac{\lambda}{\lambda _J} \right)  ^2 \left( \rho - 1 \right) \\
    \end{array}
\end{equation}
where we have dropped all the bars for the sake of simplicity.

\noindent At this stage we introduce the stretched coordinates for reductive perturbation technique \cite{rpt} :
\begin{equation}\label{rpt}
    \xi = \displaystyle \sqrt{\epsilon} \left( x - V t \right) \ , \ \tau = \epsilon ^{3/2} t
\end{equation}
where $\epsilon$ is a small non zero parameter proportional to the amplitude of perturbation, and $V$ is the Mach number. The dynamical variables $\rho$ and $v$ are expanded about their equilibrium value, in a power series of $\epsilon$, as
\begin{equation}\label{v-rho}
    v = \epsilon v ^{(1)} + \epsilon ^2 v ^{(2)} + \cdots \ , \qquad \rho = 1 + \epsilon \rho ^{(1)} + \epsilon ^2 \rho ^{(2)} + \cdots
\end{equation}
subject to the boundary conditions that all perturbed values and their derivatives vanish at $ \xi = - \infty$ :
\begin{equation}\label{bc}
    \displaystyle \rho ^{(1)} , \ v^{(1)}  \ \rightarrow \ 0 \ \rm{as} \ \xi \ \rightarrow \ - \infty
\end{equation}
Instead of going into detailed algebraic calculations, we shall just quote the results here.
The zeroth order terms of equations (\ref{cont-2}) and (\ref{mom-2}) yield $ v^{(1)} = V \rho ^{(1)}$ and $ \rho ^{(1)} = V v^{(1)} $ so that $V=1$ and
\begin{equation}\label{v1}
    v^{(1)} = \rho ^{(1)}
\end{equation}
For the perturbation expansion to be valid with the inclusion of gravitational and dissipative terms, we must have the following scaling :
$$\nu = \displaystyle \nu _0 \epsilon ^{3/2} \ , \ \displaystyle \frac{\lambda }{\lambda _J } \sim 0(\epsilon) \ \Rightarrow  \ \left(\frac{\lambda }{\lambda _J } \right) ^2 \sim 0(\epsilon ^2)$$
This is in agreement with our assumption for the stable case $k_J \ll k$; i.e., perturbation wavelength $\lambda \ll $ Jeans wavelength $\lambda _J$.
After some lengthy but straightforward algebra, we obtain the final equation as
\begin{equation}\label{final}
\begin{array}{lcl}
    & & \displaystyle \frac{\partial}{\partial \xi } \left[ \frac{\partial \rho ^{(1)}}{\partial \tau}
    + \frac{3}{2} \rho ^{(1)} \frac{\partial \rho ^{(1)}}{\partial \xi}
    - \frac{H_Q ^2}{2} \frac{\partial ^3 \rho ^{(1)}}{\partial \xi ^3} + \frac{\nu _0}{2 H_Q} \rho ^{(1)}  \right] \\ \\
    & & \displaystyle + \mu \rho ^{(1)} = 0 \\
\end{array}
\end{equation}
where $ \mu = 2 \pi ^2 \displaystyle \left( \frac{\lambda }{\lambda _J } \right) ^2 $. \\
Integrating once wrt $\xi$, equation (\ref{final}) takes the form
\begin{equation}\label{int}
\begin{array}{lcl}
    & & \displaystyle \frac{\partial \rho ^{(1)}}{\partial \tau}
    + \frac{3}{2} \rho ^{(1)} \frac{\partial \rho ^{(1)}}{\partial \xi}
    - \frac{H_Q ^2}{2} \frac{\partial ^3 \rho ^{(1)}}{\partial \xi ^3} + \frac{\nu _0}{2 H_Q} \rho ^{(1)} \\ \\
    & & \displaystyle + \mu \int _{- \infty} ^{\xi} \rho ^{(1)} d \xi ^{\prime} = 0 \\
    \end{array}
\end{equation}
Thus the final equation is obtained as a modified form of KdV equation,
with the quantum term (proportional to $H_Q$) playing the role of dispersion.
Our next step would be to study the effect of dissipation on the nonlinear structures.

In the absence of both gravitational and dissipative effects ($\mu = 0, \ \nu _0 = 0$),
we get back the KdV equation, with negative dispersion. Incidentally, the sign of dispersion (positive or negative)
is important in determining only the direction of propagation of the wave. To obtain the soliton energy, we
multiply eq. (\ref{int}) by $\rho ^{(1)}$, and integrate the same,
with proper boundary conditions; viz., $\psi $ and its derivatives vanish at $\xi \rightarrow \pm \infty$.
In the absence of dissipative and gravitational effects ($\nu _0 = 0, \ \mu = 0$),
the soliton energy, given by
\begin{equation}\label{energy}
   I = \displaystyle \frac{1}{2} \int_{- \infty} ^{\infty} \rho ^{(1) ^ 2} (\xi, \tau) d \xi
\end{equation}
is conserved : $\displaystyle \partial I / \partial \tau = 0 $. The soliton solution is obtained as
\begin{equation}\label{sol}
    \displaystyle \rho ^{(1) } (\xi, \tau) = - N \ {\rm{sech}} ^2 \left( \frac{\xi + U \tau}{W} \right)
\end{equation}
where $U$ denotes the soliton velocity, $N$ is the dimensionless soliton
amplitude, and $W$ is its dimensionless spatial width, inter-related through the relation
\begin{equation}\label{unw}
    U = \displaystyle \frac{N}{2} \ , \ W = \displaystyle \frac{2 H_Q}{\sqrt{N}}
\end{equation}
Thus as the velocity of the soliton increases, its width
decreases. These observations are consistent with those mentioned
in \cite{samiran-pre84}. Substituting (\ref{sol}) in
(\ref{energy}), the soliton energy in the absence of gravitational
and dissipative effects is obtained as $I = \displaystyle
\frac{2}{3} N^2 W = \frac{4}{3} N^{3/2} H_Q $ .

However, in the presence of gravitational and dissipative effects, eq. (\ref{int}) does not
represent a completely integrable Hamiltonian. The soliton energy is no longer conserved :
$     \displaystyle \partial I / \partial \tau \neq 0 $. In fact,
\begin{equation}\label{e-dis}
    \displaystyle \frac{\partial I}{\partial \tau} = - \frac{\nu _0}{H_Q} I - \mu \int _{- \infty} ^{\infty}
    \rho ^{(1)} (\xi, \tau) \left( \int _{- \infty} ^ {\xi} \rho ^{(1)} (\xi ^{\prime} , \tau) \ d \xi ^{\prime} \right) d \xi
    \end{equation}
Since $\nu _0$ and $\mu$ are very small parameters compared to unity, we can assume a slowly varying time-dependent form
of the solution in (\ref{sol}):
\begin{equation}\label{sol-t}
    \displaystyle \rho ^{(1) } (\xi, \tau) = - N (\tau) \ {\rm{sech}} ^2 \left( \frac{\xi + U (\tau) \tau}{W (\tau)} \right)
\end{equation}
Using (\ref{sol-t}) in (\ref{energy}), the soliton energy turns out to be
\begin{equation}\label{e-t}
     I(\tau) = \displaystyle \frac{4}{3} [N(\tau)] ^{3/2} H_Q
\end{equation}
Now, the integration in equation (\ref{e-dis}) can be performed
analytically, using equations (\ref{sol-t}) and (\ref{e-t}).
Finally, equation (\ref{e-dis}) reduces to
\begin{equation}\label{dn-dt}
    \displaystyle  \frac{\partial [N(\tau)] ^{3/2}}{\partial \tau}
    + \frac{\nu_0}{H_Q} [N(\tau)] ^{3/2} = - 6 \mu  H_Q N(\tau)
\end{equation}
where we have used eq. (\ref{unw}). This last equation
(\ref{dn-dt}) can be solved analytically. The final solution is
obtained as
\begin{equation}\label{n-final}
    \sqrt{N(\tau)} =  \displaystyle \sqrt{N_0} \ \left\{ \frac{1 + \tau _0}{\tau _0} \right\}
    \ e^{- \frac{\nu _0}{3 H_Q}  \tau }
\end{equation}
where $N_0$ is the initial soliton amplitude, and
\begin{equation}\label{tau-0}
    \tau _0 = \displaystyle \frac{3 \mu }{\nu _0} H_Q W(0)
\end{equation}
$W(0) = 2 H_Q / \sqrt{N_0} $ being the initial soliton width.

\begin{figure}
 \begin{center}
 \includegraphics[width = 6 cm]{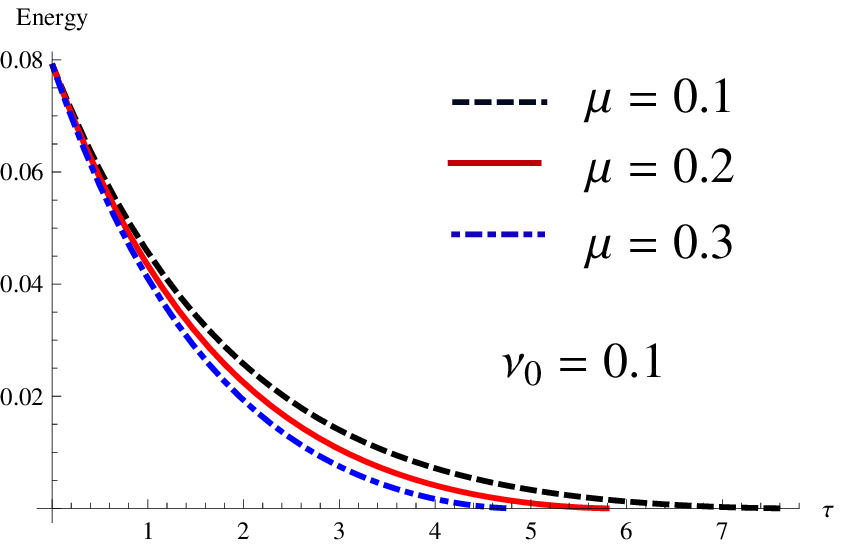} ~~~ \includegraphics[width = 6 cm]{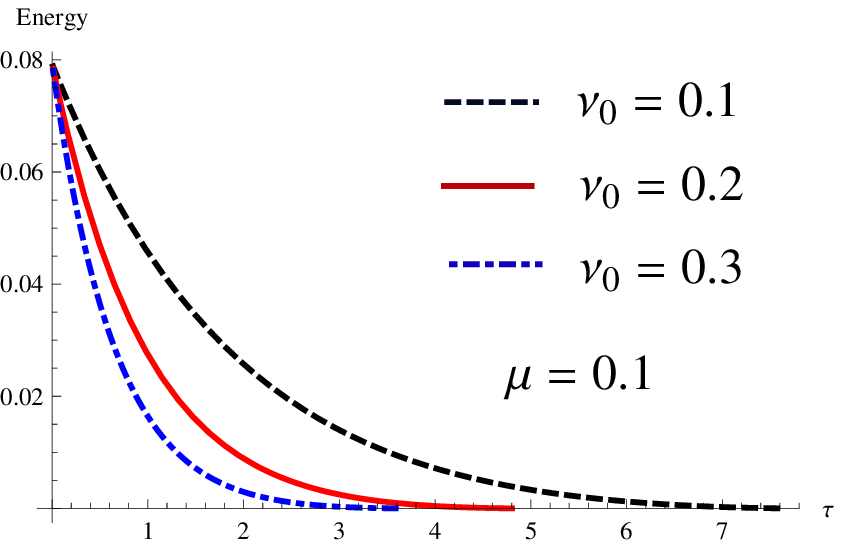} \\
  \caption{Plot of Energy $I$ against $\tau$, for different values of the parameters $\mu$ and $\nu _0$. }
 \end{center}
 \end{figure}

\begin{figure}
 \begin{center}
 \includegraphics[width = 6 cm]{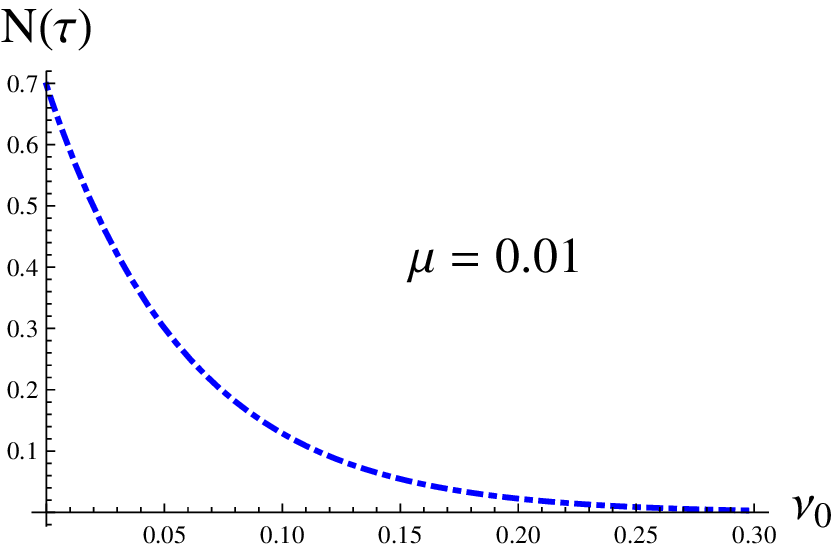} ~~~ \includegraphics[width = 6 cm]{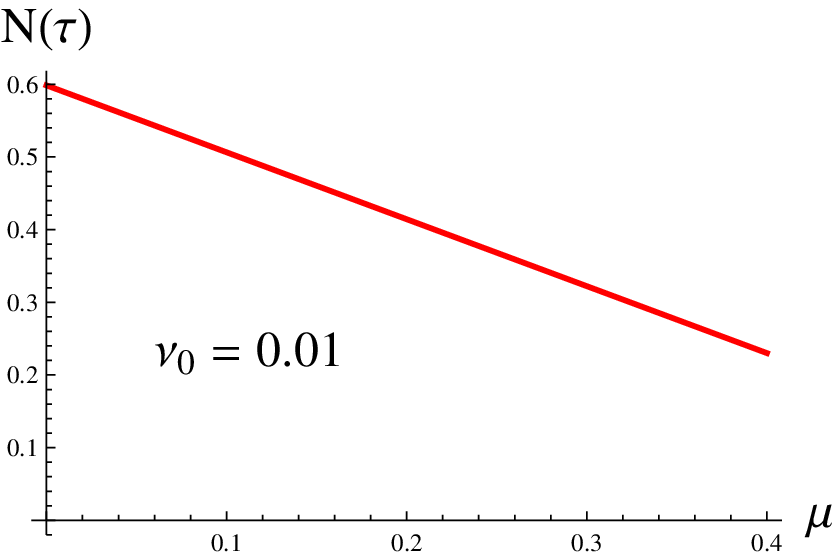} ~~~
 \includegraphics[width = 6 cm]{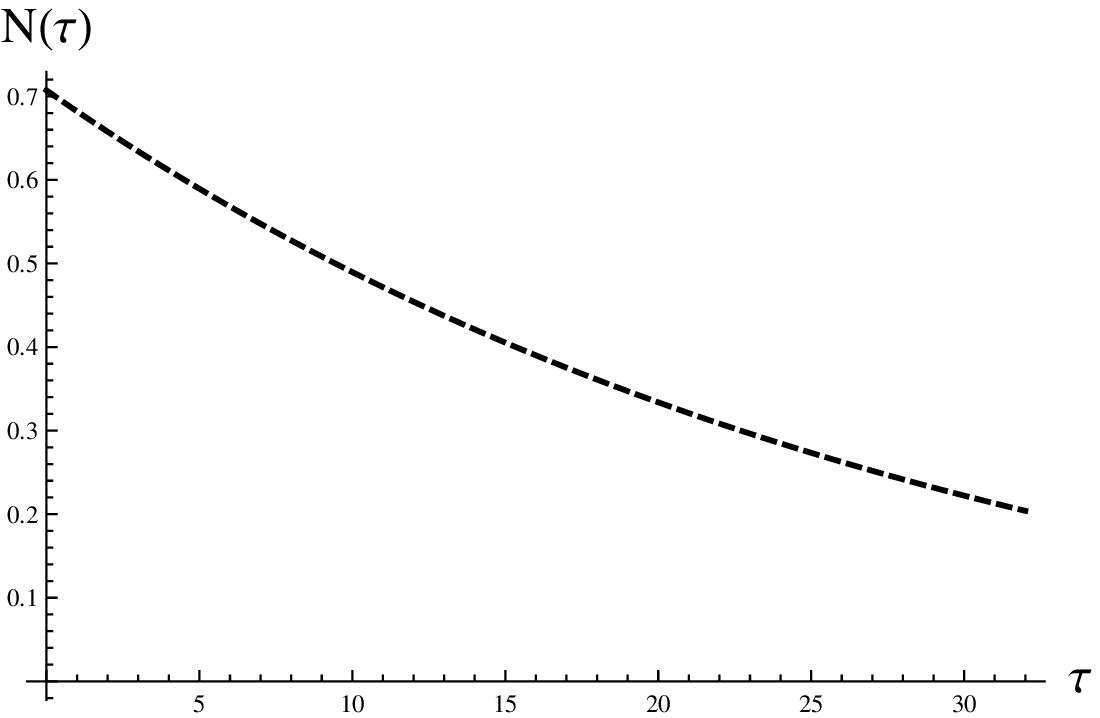} \\
  \caption{Plots showing variation of the amplitude $ N(\tau)$ w.r.t. various parameters --- viz.,
  $\mu$, $\nu _0$ and $\tau $. }
 \end{center}
 \end{figure}

Thus the amplitude of the soliton decreases with time $\tau$. This
indicates that the soliton exists for a finite time only --- say
$\tau _c$, given by
\begin{equation}\label{tau-c}
  \displaystyle \tau _c = \frac{3 H_Q}{\nu _0} \ \ln \left[ \frac{1 + \tau _0}{\tau _0} \right]
\end{equation}
To the best of our knowledge, equations (\ref{n-final}) and
(\ref{tau-c}) are new results, not reported elsewhere.

Thus the soliton exists for a finite time only --- so long as the
soliton energy is positive. We plot the variation of the soliton
energy in Fig. 2 and the soliton amplitude in Fig. 3, with $\tau$,
for different values of the dissipation factor $\nu$ and the
gravitational parameter $\mu$. Both figures indicate that the
effect of dissipation is far more pronounced than the effect of
gravitation.

Fig. 4 gives the 3-dim. plot of the soliton solution $\rho ^{(1)}$
against $\xi$ and $\tau$. Clearly, the soliton amplitude decreases
and its width increases, such that the soliton vanishes after a
finite time. This factor is more evident in the 2-dim. plot of the
soliton solution against $\xi$ in Fig. 5, for a fixed time $\tau$,
for various values of the dissipation parameter $\nu$.

\begin{figure}
 \begin{center}
 \includegraphics[width = 7 cm]{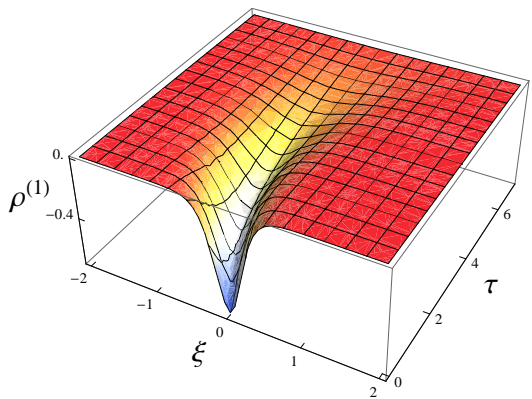}
  \caption{3-dim. plot of the solution $\rho ^{(1)}$ against $\xi$ and $\tau$, for $\mu = 0.1, \ \nu _0 = 0.04 $.
  The amplitude diminishes with progressing $\tau$.}
 \end{center}
 \end{figure}

\begin{figure}
 \begin{center}
 \includegraphics[width = 7 cm]{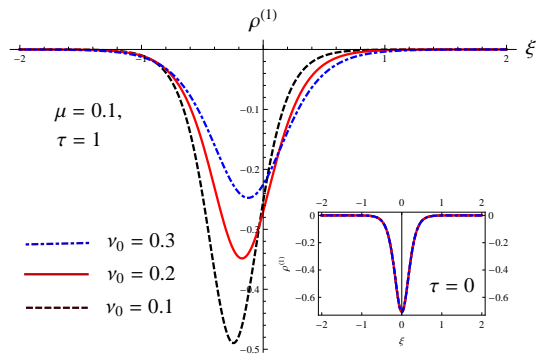}
  \caption{Plot of the solution $\rho ^{(1)}$ against $\xi$, at $\tau = 1$, for $\mu = 0.1$.
  The amplitude diminishes and width increases with increasing $\nu _0$, for $\tau \neq 0$.
  Inset plot shows the solution has same amplitude and width for different $\nu _0$ at $\tau = 0$.}
 \end{center}
 \end{figure}

\section{5 Conclusions and Discussions}

In this work we have performed an analytical study of nonlinear wave propagation in a self gravitating
Bose Einstein condensate \textbf{in 3 dimensions}, at ultra cold temperatures, in the framework of Gross-Pitaevskii equation, in the presence of dissipation. Though the mechanism of dissipation is not very well understood, nevertheless,
experimental results prove the presence of a dissipative term. The linear dispersion relation shows that the effect of dissipation is to
introduce an imaginary component to $\omega $, which damps the nonlinear wave. This is shown graphically in
Fig. 1. In the absence of dissipation, the system is stable only
if the Jeans wave number is less than the perturbation wave number : $k_J < k$. This strict condition gets relaxed in the presence of dissipation.
Thus dissipation tries to suppress dynamical instability.

\textbf{To investigate the evolution of small nonlinear
structures, we considered the 3-dimensional Bose gas to be confined tightly in two spatial dimensions, and weakly
propagating in the x direction only. This is possible at very low temperatures, when longitudinal dynamics dominate over the transverse one. The analysis was carried out} using reductive perturbation technique, for the stable case --- perturbation wavelength
$\lambda $ less than the Jeans wavelength $\lambda _J$. The final form was obtained as a modified form of the KdV equation,
giving soliton solution in some finite time interval. The soliton energy, velocity and amplitude were observed to decay with time. However,
the width of the soliton increases such that the product of the soliton amplitude and square of the width remains constant.
Though gravitational effect $\mu$ and dissipative term $\nu$ have similar effects on the propagation of nonlinear structures,
the effect of dissipation is far more pronounced than that of gravitation. Figures 2 to 5 consolidate our observations. The graphs were plotted
for physically relevant parameters mentioned in Section 2 \cite{physical}. \textbf{We propose to extend the evolution of small amplitude weak nonlinear
structures, in two and three spatial dimensions in the near future. }

With a surge of activity in theoretical and experimental studies related to Bose Einstein condensates in recent times,  our present work may be of significant relevance in this field.

\section{Acknowledgement} \textbf{The authors thank the unknown referees for their valuable comments.} One of the authors, AS, thanks the Department of Science and Technology,
Government of India, for financial support, through its grant SR/WOS-A/PM-14/2016.

\end{document}